\documentclass[aip, floatfix, amsmath, amssymb, reprint]{revtex4-1}
\usepackage{graphicx}
\usepackage{dcolumn}
\usepackage{bm}
\usepackage{color}
\usepackage{subfigure}
\usepackage{booktabs}
\usepackage{amsmath}
\usepackage{verbatim}
\usepackage{lineno}
\usepackage{xcolor} 

\newcommand{\beq}{\begin{eqnarray}}
\newcommand{\eeq}{\end{eqnarray}}

\usepackage[normalem]{ulem}

\usepackage[utf8]{inputenc}
\usepackage[T1]{fontenc}
\usepackage{mathptmx}
\usepackage{romannum}
\usepackage[normalem]{ulem}

\begin{document}

\preprint{AIP/123-QED}

\title[]{Revisiting the Thickness of the Air-Water Interface from Two Extremes of Interface Hydrogen Bond Dynamics}

\author{Gang Huang}
\affiliation{
Institute of Theoretical Physics, Chinese Academy of Sciences, Beijing 100190, China
}
\author{Jie Huang*}
\email{jie.huang@aalto.fi}
\affiliation{
Department of Applied Physics, Aalto University, Helsinki FI-00076, Finland
}

\begin{abstract}
The air-water interface plays a crucial role in many aspects of science, because of its unique properties, such as a two-dimensional hydrogen bond (HB) network and completely different HB dynamics compared to bulk water. However, accurately determining the boundary of interfacial and bulk water, that is, the thickness of the air-water interface, still challenges experimentalists. Various simulation-based methods have been developed to estimate the thickness, converging on a range of approximately 3--10 (\AA). In this study, we introduce a novel approach, grounded in density functional theory-based molecular dynamics and deep potential molecular dynamics simulations, to measure the air-water interface thickness, offering a different perspective based on prior research. To capture realistic HB dynamics in the air-water interface, two extreme scenarios of the interface HB dynamics are obtained: one {underestimates} the interface HB dynamics, while the other {overestimates} it. Surprisingly, our results suggest that the interface HB dynamics in both scenarios converges as the thickness of the air-water interface increases to 4 (\AA). This convergence point, indicative of the realistic interface thickness, is also validated by our calculation of anisotropic decay of OH stretch and the free OH dynamics at the air-water interface.
\end{abstract}

\maketitle

\section{Introduction}
\label{sec:intro}
The air-water interface has been the subject of extensive study due to its ubiquity in nature and its unusual macroscopic properties as a model system for aqueous hydrophobic interfaces~\cite{Morita2000, Jungwirth2001, Wilson2002, Raymond2003, ShenYR2006, Smith2007, Sovago2008, Sovago2009, Stiopkin2011, Hsieh2011, Nihonyanagi2011, Singh2014, Nihonyanagi2015, Ishiyama2021, Ahmed2022, Omranpour2024perspective}. It is widely accepted that water molecules behave in a completely different manner at the interface than in the bulk phase~\cite{LeeCY1984, DangLX1997, Hsieh2013}. 

Advances in the study of hydrogen bond (HB) dynamics at the air-water interface have been significant. Liu et al.~\cite{LiuPu2005} used molecular dynamics (MD) simulations to demonstrate faster HB breaking and forming at the interface than bulk water, attributed to quicker translational diffusion. {From sum frequency generation (SFG) vibrational spectroscopy, Gan et al.~\cite{Gan2006} found that at the air-water interface, singly hydrogen (H)-bonded water molecules align almost parallel to the interface with limited orientational variation, while doubly H-bonded donor molecules orient their dipole vectors away from the liquid phase, highlighting diverse behaviors among interfacial water molecules.} Almost concurrently, through time-resolved SFG vibrational spectroscopy, McGuire and Shen\cite{Mcguire2006} observed ultrafast vibrational dynamics at the interface, noting that the relaxation behaviors of interfacially bonded OH stretch modes on subpicosecond time scales were akin to those in bulk water, encompassing spectral diffusion, vibrational relaxation, and thermalization.  Pioneering work by Tahara's group~\cite{Singh2013, Inoue2016}, which presented the first two-dimensional heterodyne-detected vibrational SFG (2D HD-VSFG) spectra of the OH stretch region at the interface, 
highlighted diverse behaviors among interfacial HB OH groups. Subsequent studies, including those by Jeon et al.~\cite{Jeon2017} and Ojha and K{\"u}hne~\cite{Ojha2021}, have used MD and \emph{ab initio} MD (AIMD) simulations to explore the structure and dynamics of interfacial water, uncovering weaker H-bonds and faster vibrational spectral dynamics of free OH groups compared to H-bonded OH groups at the interface. These collective insights enhance our understanding of the vibrational energy relaxation, HB dynamics, and interactions of water molecules at the air-water interface. Building upon this knowledge of interfacial behavior, significant efforts have also been directed toward quantifying the physical characteristics of the interface, including its thickness, which plays a crucial role in understanding its molecular interactions and behavior.

The air-water interface thickness has been measured via ellipsometry~\cite{Raman1927, McBain1939, Kinosita1965}, relative permittivity measurements~\cite{Fumagalli2018},
X-ray reflectivity~\cite{Braslau1985, Braslau1988},
SFG spectroscopy~\cite{RS91, Du93, VO05, Groenzin2007Single-crystal, Shen2012Basic, Ishiyama2012Origin, Shen2013, Singh2013, Ohto2015, TangFJ2020}, classical MD simulations\cite{LiuPu2005, Kuo2006, Wick2007, Morita_2008, Sedlmeier2009, FanYB2009, Ishiyama2009, Willard2010, VilaVerde2012, Ishiyama2015, Nagata2015} and the AIMD simulations\cite{Kuehne2011, Nagata2012, Kessler2015, Jeon2017, Ohto2019, TangFJ2020} to mention just a few. There is a consensus that the thickness of the air-water interface is about 3--10 (\AA)~\cite{PJ06, Sedlmeier2009, Ishiyama2009, Kuo2006, Wick2007, FanYB2009, Kuehne2011, Baer2014, Kessler2015, Nguyen2020, Peng2021}.
Nonetheless, accurately determining the thickness remains experimentally challenging. The MD and Monte Carlo (MC) simulations of the air-water interface yield molecular-level information not readily available in experiments. These simulations, which utilize various intermolecular potential functions, have played a crucial role in estimating the thickness~\cite{Townsend1985, Wilson1987, Matsumoto1988, Townsend1991, Lie1993, Alejandre1995, Taylor1996, Dang1997, TangFJ2020}. Additionally, density functional theory-based MD (DFTMD) simulations~\cite{CP, Marx2000, RC02, Kuo2004b, Kuehne2007, Kuehne2011, Khaliullin2013} also offer a predictive platform for understanding density profiles and determining the thickness of the air-liquid interfaces~\cite{Blas2001, Kuehne2011, Kessler2015, Pezzotti2017}. Nevertheless, DFTMD simulations are constrained by limitations in time and the number of molecules they can model. Traditional force field approaches, however, often lack the accuracy required to describe complex interface systems. \cite{Schran2021} Recently, deep potential molecular dynamics (DeePMD) simulations based on machine learning potential (MLPs) have emerged as a promising alternative, offering a solution to the accuracy-versus-efficiency dilemma in molecular simulations. \cite{Zeng2023} One of the most accurate MLPs for water is MB-pol, which accurately reproduces many properties of water across the phase diagram. \cite{Babin2013, Babin2014, Medders2014, Bore2023, Muniz2021} Moreover, MB-pol has been used to obtain the VSFG and surface tension of air-water interface, demonstrating excellent agreement between theoretical predictions and experimental measurements. \cite{Medders_2016} 

Inspired by the above experimental and simulation results, and with the motivation of capturing realistic HB dynamics at interfaces, we have designed an approach based on two extreme scenarios of interface HB dynamics by utilizing the trajectories of DFTMD and DeePMD simulations based on MB-pol. In the first scenario, for the set of molecules located in the interface layer at given sampling times, we use the Luzar-Chandler (LC) HB population operator~\cite{LC96} to obtain the HB dynamics of these interface molecules. In the second scenario, taking inspiration from Luzar and Chandler's HB population and the characteristic function introduced by Giberti and Hassanali~\cite{Giberti2017},  we have developed an interface HB (IHB) population operator. This operator aims to provide a refined understanding of HB dynamics specifically at the interface.

The Luzar-Chandler HB population is utilized to describe whether a pair of labeled molecules form H-bonds. And the characteristic function developed by Giberti and Hassanali describes whether a specific molecule belongs to the interface region. Therefore, our newly defined interface HB population can describe whether a labeled pair of molecules is within the interface \emph{and} connected by H-bonds at any given moment. This dual condition provides a more detailed understanding of interfacial HB dynamics. 

Due to the thermal motion of water molecules,  two scenarios may occur. In the first scenario, the water molecules under observation might transition into the bulk phase. In the second scenario, if a water molecule resides in the interface region \emph{and} its H-bonded partner moves outside the interface area, then such a pair of molecules will no longer be of concern.
Based on the study of interfacial water dynamics by Liu et al.~\cite{LiuPu2005}, Gan et al. \cite{Gan2006}, Singh et al.~\cite{Singh2013} and Jeon et al.~\cite{Jeon2017}, as well as the investigation into the time-dependent spectral evolution of H-bonded and free water molecules by Ojha and K\"uhne~\cite{Ojha2021}, our approach indicates that the HB dynamics derived from the first scenario will be slower the genuine interfacial HB dynamics. Conversely, the one obtained from the second scenario will exhibit faster dynamics than the genuine one.
 
Building upon the foundation laid by methods reliant on the density criterion~\cite{Townsend1991, Beaglehole1993, Lie1993, Taylor1996, Sokhan1997, Alejandre1995, Morita2000, Paul2004, Morita2006, ZhaoJin2018}, our approach introduces an alternative way of determining interface thickness through the analysis of the convergence of interfacial HB dynamics properties. This approach effectively bypasses the necessity of accounting for liquid density. As such, it offers another perspective for measuring the thickness of the air-water interface. Furthermore, the principles underlying our approach hold potential for application to a broader range of systems, such as solution interfaces and ion shells, offering a flexible tool for interface studies. 

\section{Methods}
\label{sec:hbd_interface}
\begin{figure}[!ht]
    \centering
    \includegraphics [width=0.4\textwidth] {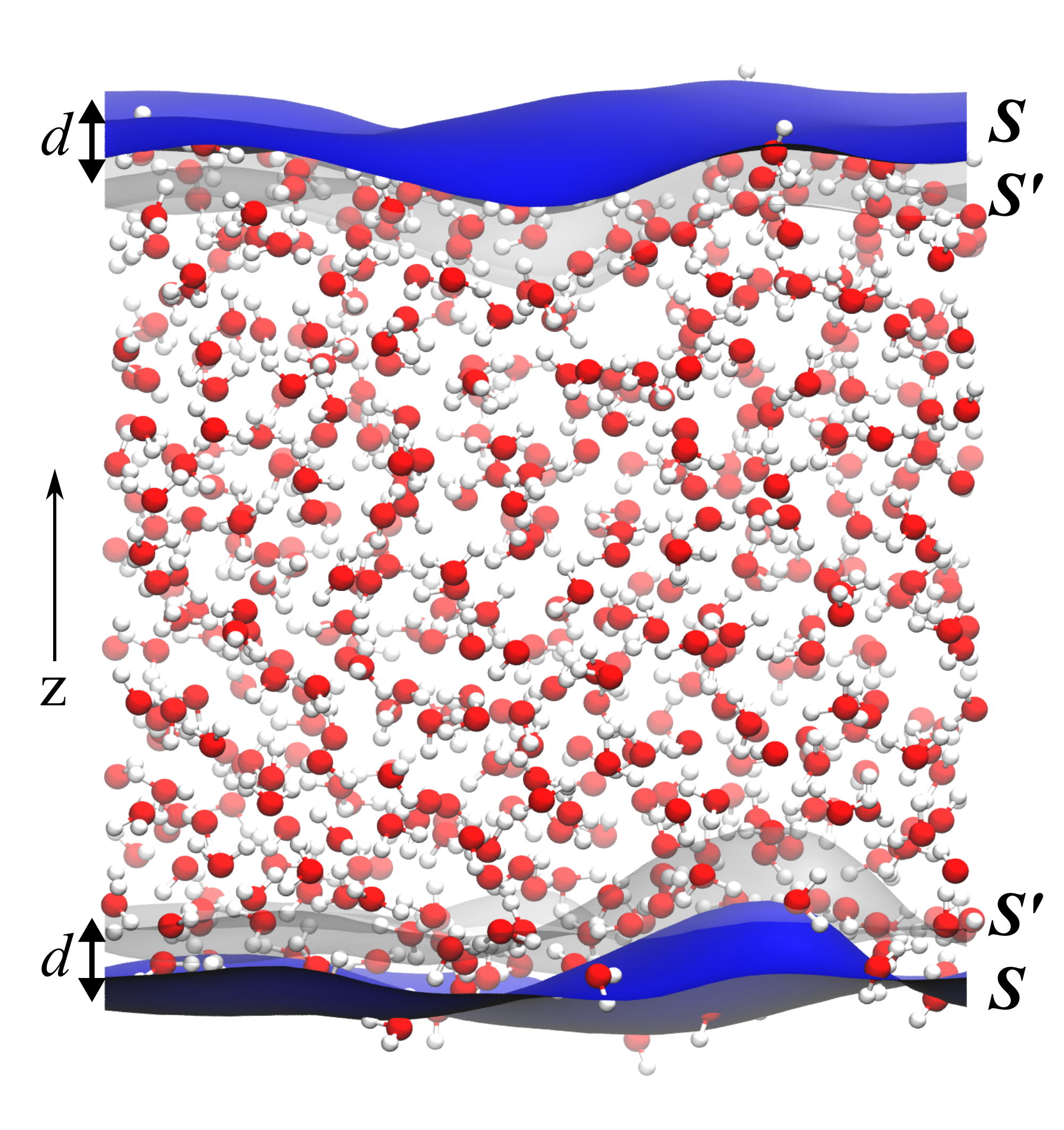}
    \setlength{\abovecaptionskip}{0pt}
    \caption{
      Slab of water containing 512 water molecules with the instantaneous surface $\mathcal{S}$ represented as a blue mesh on the upper and lower phase boundary. Grey surface, which represents an imaginary interface $\mathcal{S}'$, is obtained by translating the surface to the inside of the system along the normal ($z$-axis) by distance  $d$. This variable $d$ is then utilized to measure the thickness of the air-water interface.}
    \label{fig:Fig1}
\end{figure}

Due to molecular motions, the identity of molecules at the interface changes with time, and generally useful procedures for identifying interfaces must accommodate these motions. The air-water boundary is modeled with the Willard-Chandler instantaneous surface~\cite{Willard2010, Pezzotti2017, Serva2018}. Figure ~\ref{fig:Fig1}
illustrates the obtained interfaces for one configuration of a slab of water.  

For the slab in the cuboid simulation box,  an imaginary surface $\mathcal{S}'(t)$ is obtained by translating the surface $\mathcal{S}(t)$ along the system's normal (into bulk) to a distance $d$.  The region between the two surfaces $\mathcal{S}(t)$ and $\mathcal{S}'(t)$ is defined as \emph{the air-water interface}. 
 {Below we will combine two extreme scenarios to investigate the HB dynamics at the instantaneous air-water interface.}

\subsection{Scenario 1: HB Dynamics Based on the Luzar-Chandler HB Population}

 {As the first scenario we use the Luzar-Chandler HB population and employ a technique that samples water molecules right at the instantaneous interface for certain sampling time points. 
 In Scenario 1, we divide the simulation trajectory into multiple subtrajectories of length $t_{\rm traj}$.
Within these subtrajectories, the majority of water molecules exhibit thermal fluctuations near their equilibrium positions. 
Meanwhile, a select group of molecules initially at the interface may transition into the bulk phase at a later time. Due to the inclusion of H-bonds in the bulk phase, this scenario tends to \emph{underestimate} the breaking rate of the H-bonds at the interface. In Section 3 of this paper, we will see this result, combined with the outcome from Scenario 2, can be used to estimate the thickness of the air-water interface. This method includes three steps as follows:}

\textit{(a) Sub-trajectories.}  Sub-trajectories with a specific length of time, $t_\text{traj}$,  are selected.  In this work, $t_{\rm traj}$ is set to be $40$ (ps), which is long enough to observe HB dynamics but short enough that not all the molecules complete their transition across the interface~\cite{Madarasz2007}.

\textit{(b) Sampling.} For each time step, we identify a pair of the air-water interfaces of a specified thickness $d$ as shown in Figure~\ref{fig:Fig1}. At evenly spaced moments within $t_\text{traj}$, we select the water molecules within the interfaces. For the union of the interfacial water molecules picked at all these time moments, the Luzar-Chandler HB population-based correlation functions~\cite{LC96} across the sub-trajectory are calculated.

\textit{(c)  Statistics.} The average correlation functions across all sub-trajectories are calculated.

In this scenario, the computational procedure for calculating HB dynamics follows the same methodology as that used in works based on the Luzar-Chandler's method (LC method)\cite{Luzar2000, Benjamin2005, LiuPu2005}.  For details on this method, please refer to the Supporting Information.

\subsection{Scenario 2: HB Dynamics Based on Interface HB Population}
\label{sec:ihb}
To capture the other extreme of interfacial HB dynamics, after determining the instantaneous interface, we introduce an interface HB population operator $h^{(\text{s})}[{\mathbf r}(t)]$ as follows:
It has a value of 1 when a tagged molecular pair $i, j$ are H-bonded \emph{and} both molecules are at the interface 
with a thickness $d$, and 0 otherwise:
\begin{align}
    h^{(\text{s})}[{\mathbf r}(t)]=\left\{
    \begin{array}{rcl}
 	1       &      & {{\mathbf r}_i\in \mathcal{I}(d;t),  {\mathbf r}_j\in \mathcal{I}(d;t)}, \\  
		       &      & {j\in \mathcal{B}_i(t)}; \\   \label{eqn:h_s}
		0       &      & {\text{otherwise}}
    \end{array} \right.
\end{align}
 {where ${\mathbf r}(t)$ is the configuration of the system at time $t$, ${\mathbf r}_i$ is the position coordinate of the oxygen atom in the $i$-th water molecule, ${\mathcal{B}_i(t)}$ denotes the set of water molecules that are H-bonded with molecule $i$ at time $t$, and $\mathcal{I}(d;t)$ is the instantaneous interface layer with thickness $d$ at time $t$.
The definition of $h^{(\text{s})}$ combines the Luzar-Chandler's HB population~\cite{LC1993, LC96}  $h$ and the characteristic function introduced by Giberti and Hassanali~\cite{Giberti2017}.}
Then the correlation function $c^{(\text{s})}(t)$ that describes the fluctuation of H-bonds \emph{at the interface}:
\begin{eqnarray}
    c^{(\text{s})}(t)=\frac{\langle h^{(\text{s})}(0)h^{(\text{s})}(t) \rangle}{\langle h^{(\text{s})}\rangle}
    \label{eq:C_s_HB},
\end{eqnarray}
can be obtained. Similar to functions $n(t)$ and $k(t)$ in ref~\citenum{LC96} (eqs 1 and 2  in Supporting Information),
the corresponding correlation function 
\begin{eqnarray}
    n^{(\text{s})}(t)=\frac{\langle h^{(\text{s})}(0)[1-h^{(\text{s})}(t)]h^{(\text{d,s})} \rangle}{\langle h^{(\text{s})}\rangle}
    \label{eq:n_s_HB},
\end{eqnarray}
and interface reactive flux function
\begin{eqnarray}
    k^{(\text{s})}(t)= -\frac{\mathrm{d}c^{(\text{s})}(t)}{\mathrm{d}t} 
    \label{eq:k_s_HB}
\end{eqnarray}
are obtained.
The $h^{(\text{d,s})}(t)$ is 1 when a tagged pair of water molecules $i$, $j$ is \emph{at the interface} and the interoxygen distance between the two molecules is less than the cutoff radius  $r^{\mathrm{c}}_{\mathrm{OO}}$ at time $t$,
and 0 otherwise, i.e.,
\begin{align}
    h^{(\text{d,s})}[{\mathbf r}(t)]=\left\{
    \begin{array}{rcl}
        1       &      & {{\mathbf r}_i\in \mathcal{I}(d;t),  {\mathbf r}_j\in \mathcal{I}(d;t)}, \\  
		&      &   { |\mathbf{r}_i-\mathbf{r}_j| < r_{\text{OO}}^{\text{c}}; }\\ 
	0       &      & {\text{otherwise.}} 
    \end{array} \right.
    \label{eq:h_s}
\end{align}
Therefore, $n^{(\text{s})}(t)$ represents the probability at time $t$ that a tagged pair of initially H-bonded water molecules {at the interface} are unbonded but remain at the interface and separated by less than $r_{\text{OO}}^{\text c}$; $k^{(\text{s})}(t)$ measures the effective decay rate of H-bonds {at the interface}.
The functions defined in eqs \ref{eq:C_s_HB}--\ref{eq:k_s_HB} are used to determine the reaction rate constants of breaking and reforming and the lifetimes of H-bonds at the interface by~\cite{LC96, Khaliullin2013}
\begin{eqnarray}
    k^{\rm (s)}(t) = kc^{\rm (s)}(t)-k'n^{\rm (s)}(t).\nonumber
    \label{eq:fitting_k_k_prime}
\end{eqnarray}

%

In this IHB  {scenario}, choosing the water molecules and H-bonds at the interface is accurate.
However, for some special H-bonds, if it connects such two water molecules, one is at the interface 
and the other is in the bulk phase, the HB breaking reaction rate of such H-bonds will be increased.
Therefore, in contrast to the LC method used in Scenario 1, the IHB method used in this scenario \emph{overestimates} the HB  breaking rate constant.  

{The actual HB dynamics at the interface are expected to lie between the results obtained by the LC method and the IHB method. Consequently, by integrating these two scenarios, we can achieve a more precise characterization of HB dynamics at the interface.}

\section{Results and Discussions}
\label{sec:discussions}

In this section, we apply our two-extreme approach to two system properties: HB population autocorrelation functions and HB reaction rate constants. We then determine the thickness of the air-water interface based on each of these properties. The computational details for DFTMD and DeePMD simulations are available in the Supporting Information. The results discussed in the following sections are derived from the trajectory data of the system containing 512 water molecules, modeled by MB-pol potential. The finite size effects of the main properties discussed in this article are in Supporting Information. Similar analyses for DFTMD simulations with 128 water molecules, and other system sizes in DeePMD simulations using MB-pol, are also provided in the Supporting Information. 

\subsection{HB Population Autocorrelation Functions}
\begin{figure*}[!ht]
    \centering                                         
    \includegraphics [width=0.8\textwidth] {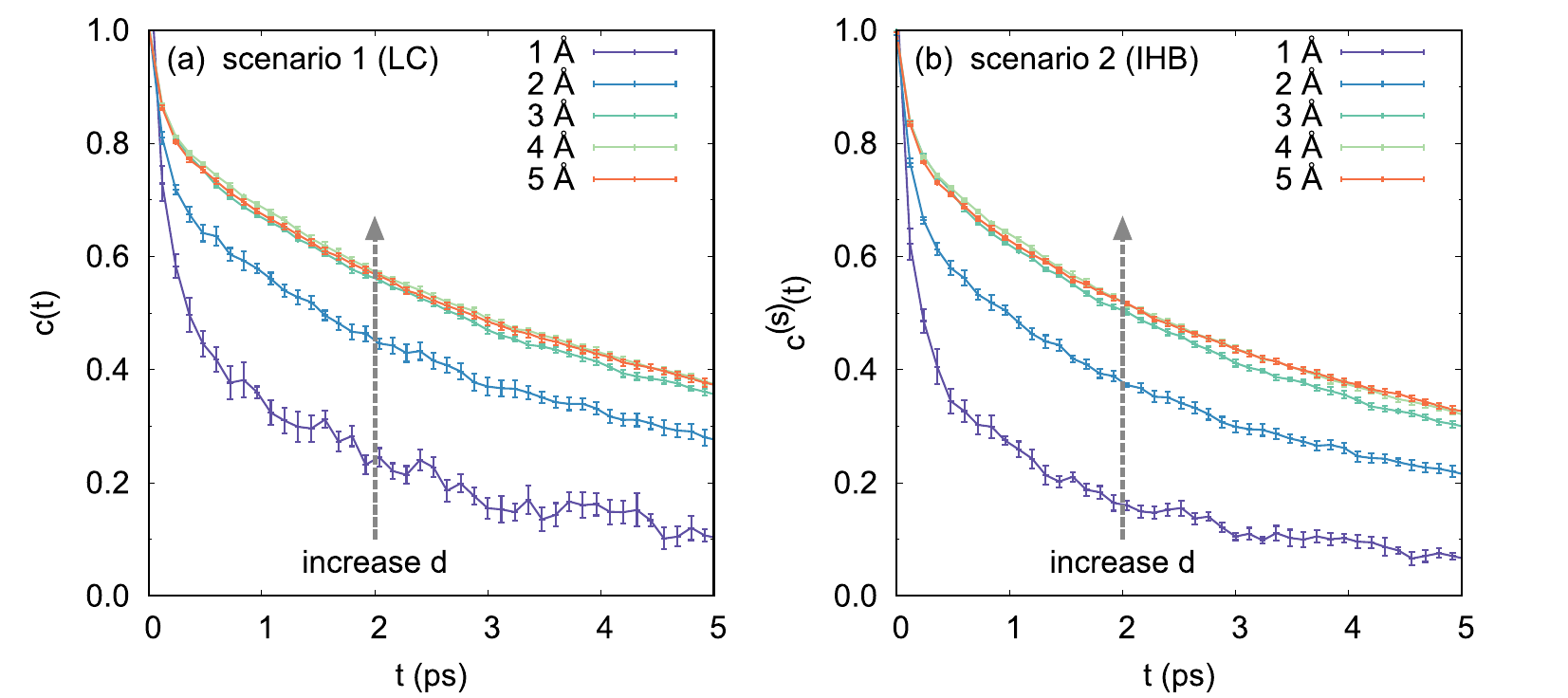}
    \setlength{\abovecaptionskip}{0pt}
    \caption{Autocorrelation functions $c(t)$ and $c^{(\text{s})}(t)$ for interface H-bonds with different thickness $d$ for (a) Scenario 1 (LC) and (b) Scenario 2 (IHB). Two notable features emerge: (i) As $d$ increases, both $c(t)$ and $c^{\rm (s)}(t)$ eventually approach a stable function. (ii) Decay rate of $c^{(\text{s})}(t)$ in Scenario 2 is greater than that of $c(t)$ in Scenario 1. This behavior is visually represented by two dashed directed line segments, positioned identically to the graphs. Functions $c(t)$ and $c^{(\text{s})}(t)$ on a log-log scale are also plotted in Supporting Information.}
    \label{fig:Fig2}
\end{figure*}

Figure~\ref{fig:Fig2} illustrates the dynamic evolution of $c(t)$ and $c^{\rm (s)}(t)$ for various values of distance $d$. When $d$ is 4 (\AA) or greater, $c(t)$ and $c^{\rm (s)}(t)$ become largely invariant to further increases in $d$. 
This indicates that one can determine the thickness of the interface under different conditions by examining how the correlation function depends on distance $d$. Analysis of both scenarios reveals that the decay rate of the correlation function $c^{(\text{s})}(t)$ in Scenario 2 (IHB) surpasses that in Scenario 1 (LC).

\begin{figure}[!ht]
    \centering
    \vspace*{0.0cm}\hspace*{-0.0cm}\includegraphics [width=0.4\textwidth]{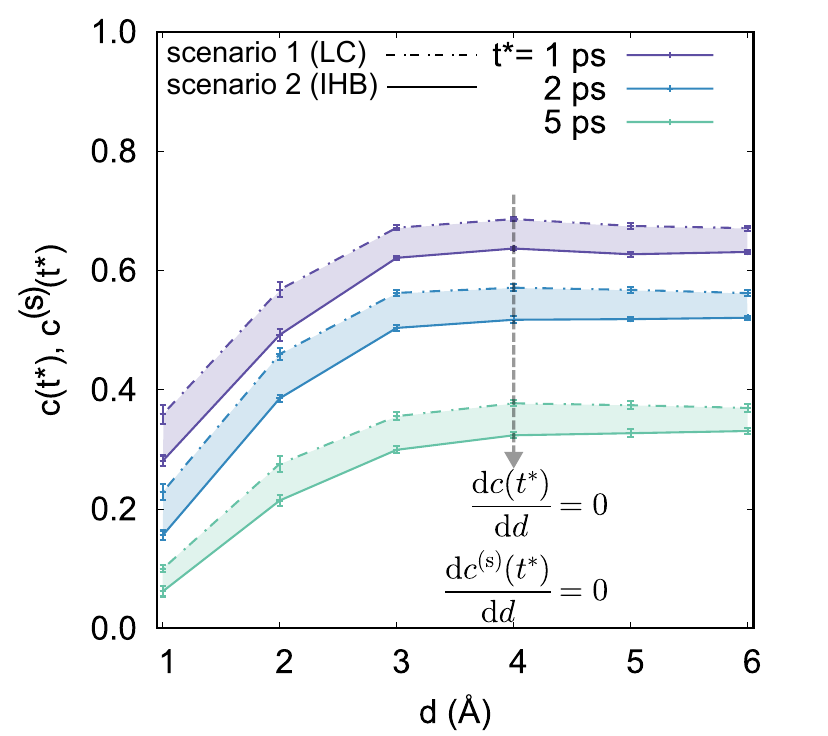}
    \setlength{\abovecaptionskip}{0pt}
    \caption{
    Dependence of the correlation functions on the distance $d$ at three reference time points $t^* = 1, 2, 5$ (ps) provides key insights into the dynamics of the air-water interface:
    (i) As $d$ increases, both $c(t^*)$ and $c^{\rm (s)}(t^*)$ exhibit an upward trend, with their rates of change gradually approaching 0. (ii) At each $t^*$, $c(t)$ is consistently slightly larger than $c^{\rm (s)}(t)$ for the same $d$. (iii) Air-water interface thickness $d_f=4 $ (\AA) is determined from the $d$-dependence of $c(t^*)$ and $c^{\rm (s)}(t^*)$, as calculated using the LC method (dot-dashed lines) and the IHB method (solid lines), respectively.}
    \label{fig:Fig3}
\end{figure}

Figure ~\ref{fig:Fig3} displays the $d$-dependence of the correlation functions at three reference time points $t^*=1,2,5$ (ps).  This figure provides a different perspective on $d$-dependence: by selecting three reference time intervals on the $t$-axis in Figure~\ref{fig:Fig2}, the values of the correlation functions, $c(t)$ and $c^{\rm (s)}(t)$, at these time intervals for each $d$ were recorded. Comparing $c(t)$ and $c^{(\text{s})}(t)$ in Figure~\ref{fig:Fig3}, $c(t)$ in Scenario 1 is always slightly larger than $c^{(\text{s})}(t)$ in Scenario 2 for the same $d$.

In Figure~\ref{fig:Fig3}, both $c(t^*)$ and $c^{\rm (s)}(t^*)$ increase as $d$ increases, eventually reaching a point where their change rates approach 0 for sufficiently large values of $d$. Since $c(t)$ and $c^{(\text{s})}(t)$ serve as the upper and lower bounds of the real interface correlation function $c^{r}(t^*)$ respectively, it logically follows that the $d$-dependence of $c^{r}(t^*)$ exhibits the same trend. Consequently, we arrive at the equations ${\rm d}C/{\rm d}d=0$ for $c(t^*)$ and $c^{\rm (s)}(t^*)$ with respect to $d$. Solving for $C=c(t^*)$ and $C=c^{\rm (s)}(t^*)$ yields solutions $d_{f1}$ and $d_{f2}$, respectively. Here $d_{f1}$ and $d_{f2}$ represent the thickness of the air-water interface as determined by the LC method and the IHB method, respectively.  The average value $d_f =(d_{f1}+d_{f2})/2$ is then obtained as the thickness of the interface for a given $t^*$.

The correlations $c(t^*)$ and $c^{\rm (s)}(t^*)$ respectively describe the relaxation characteristics of HB dynamics within the air-water interface. For both the LC and IHB methods, the values of $c(t^*)$ and $c^{\rm (s)}(t^*)$  cease to change significantly when $d$ reaches 4 (\AA), i.e., $d_{f1}\approx d_{f2}= 4$ (\AA). Thus, we determine the thickness of the air-water interface in the simulations to be $d_f=4$ (\AA). 

\subsection{HB Reaction Rate Constants}
\label{sec:rate_constants}

We further examined how the reaction rate constants of H-bonds at the air-water interface vary with $d$. For further details regarding the HB reaction rate constants, please refer to Section 2 in Supporting Information. 

\begin{figure}[!ht]
\centering
\vspace*{0.0cm}\hspace*{-0.0cm}\includegraphics [width=0.4\textwidth]{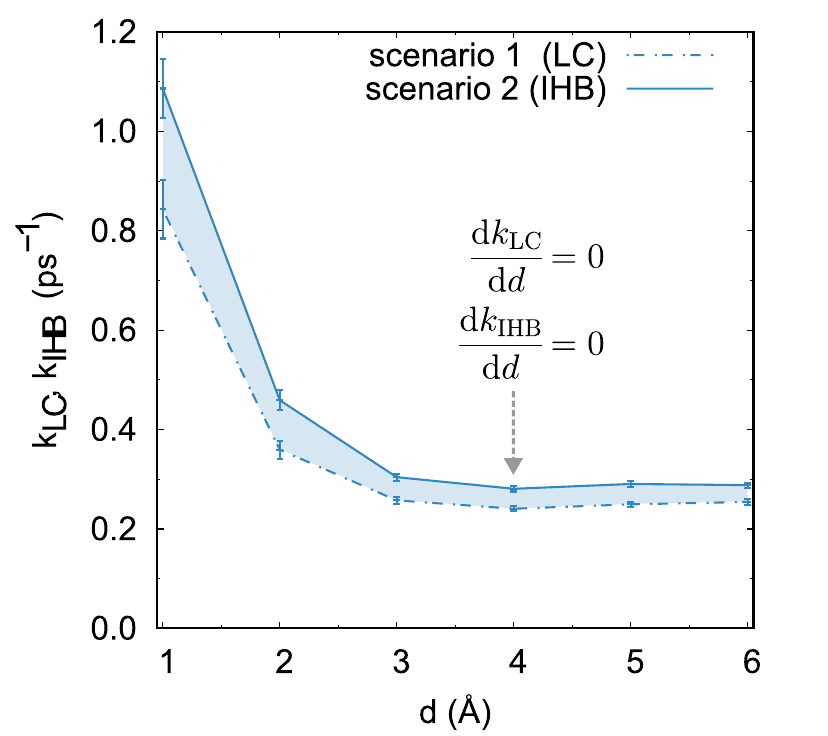}
\setlength{\abovecaptionskip}{0pt}
\caption{
Breaking HB reaction rate constants $k_{\rm LC}$ and $k_{\rm IHB}$ for the air-water interface simulated with
512 water molecules, obtained by the LC and IHB methods respectively.
(i) Both constants, $k_{\rm LC}$ and $k_{\rm IHB}$, decrease monotonically to the HB breaking rate $k_{\rm bulk}$ for the bulk water as $d$ increases.
(ii) $k_{\rm LC}$ is always smaller than $k_{\rm IHB}$ for the same $d$.
(iii) Thickness $d'_f=4$ (\AA) of the air-water interface is obtained from the $d$-dependence of the rate constant $k_{\rm LC}$ and $k_{\rm IHB}$, obtained by the LC method and the IHB method, respectively. (Consistent results for systems with 125 and 216 water molecules, please refer to the Supporting Information.)}
\label{fig:Fig4}
\end{figure}

In Figure~\ref{fig:Fig4}, we compare the breaking HB reaction rate constants\cite{LC96, Luzar2000}, $k_{\rm LC}$ and $k_{\rm IHB}$, obtained by the LC and IHB methods, respectively. 
We found that both constants, $k_{\rm LC}$ and $k_{\rm IHB}$, decrease monotonically as $d$ increases. When $d$ is large, both rate constants $k_{\rm IHB}$ and $k_{\rm LC}$ also no longer change with $d$.  
It also shows that the HB breaking rate constant $k_{\rm IHB}$ is relatively larger than $k_{\rm LC}$. This difference is related to the definitions of HB populations, $h(t)$ and $h^{\text{(s)}}(t)$. The definition of $h^{(\text{s})}(t)$ leads to an increased HB break rate at the interface. The LC method retains the original rate constant of H-bonds but may include contributions from bulk water molecules. 

Similar to the HB autocorrelation functions, we have the equations ${\rm d}K/{\rm d}d=0$, for $k_{\rm IHB}$ and $k_{\rm LC}$, with respect to $d$. 
For $K=k_{\rm IHB}$ and $K=k_{\rm LC}$, we identify solutions $d'_{f1}$ and $d'_{f2}$, respectively. Here $d'_{f1}$ and $d'_{f2}$ represent the HB reaction rate-based thickness of the interface obtained from the LC and IHB method, respectively.  We then calculate the thickness of the real air-water interface as $d'_f = (d'_{f1}+ d'_{f2})/2\approx 4$ (\AA).  This result is supported by our calculation of the $d$-dependence of the HB reforming rate constants, namely,  $k'_{\rm LC}$ and $k'_{\rm IHB}$. For more details, please refer to  Figure S7 in Supporting Information.

\subsection{Orientational Relaxation of the OH Stretch and Other Supports}

To verify our conclusion on the thickness of the air-water interface from the interfacial HB dynamics, we performed calculations on the orientational relaxation of the OH stretch from the perspective of anisotropy decay\cite{Moilanen2008}.

\begin{figure*}[!ht]%
    \centering
    \includegraphics[width=0.8\textwidth]
    {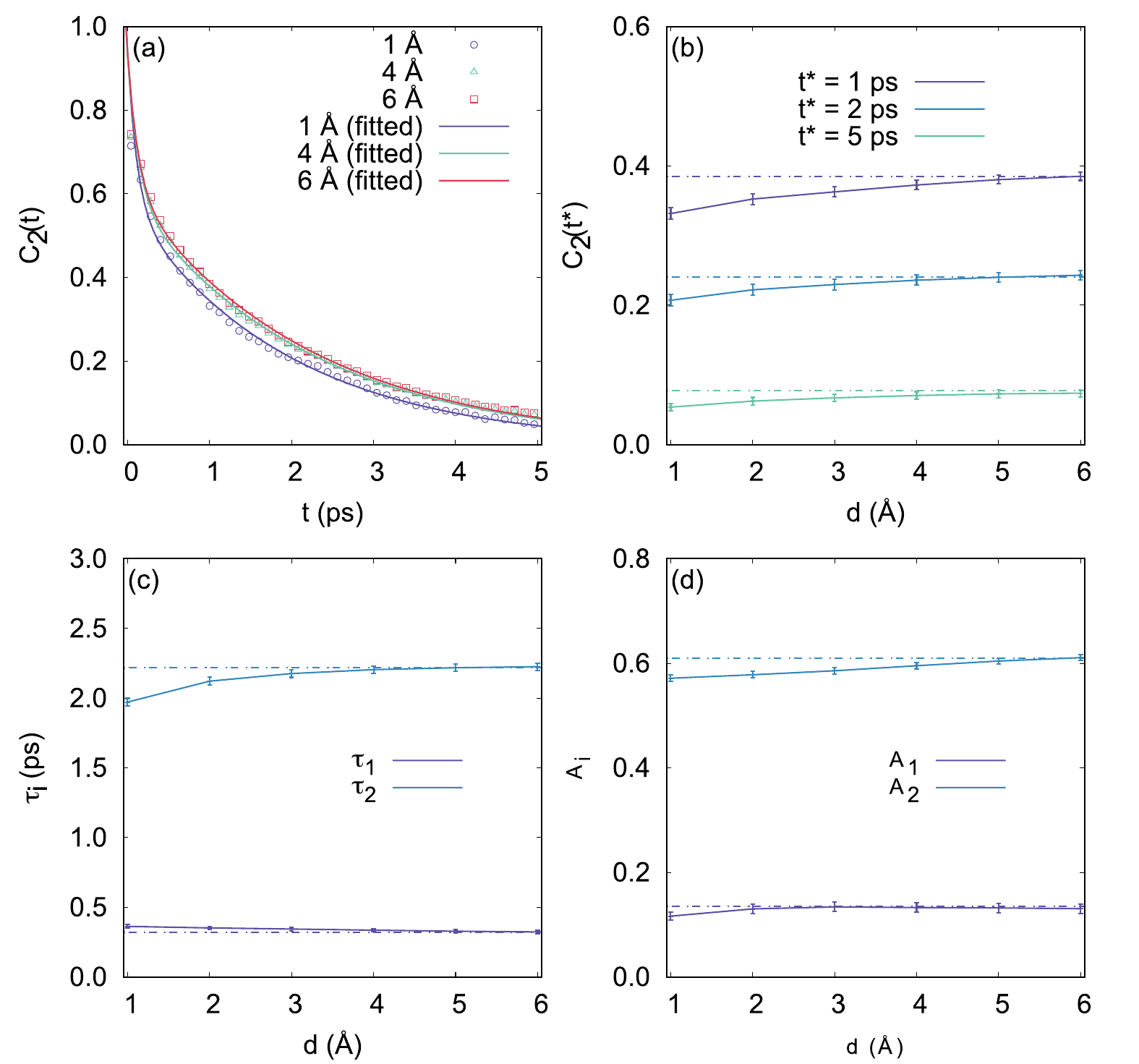}
    \caption{Orientation correlation function $C_2(t)$ for the air-water interface. (a) The $C_2(t)$ for $d=1, 4, 6$ (\AA) and the biexponential decay function  $C_2(t)= A_1 \exp(t/\tau_1) +  A_2 \exp(t/\tau_2)$. $d$-dependence of (b) $C_2(t^*)$ for water molecules at the air-water interface with $t^*= 1, 2, 5$ (ps); (c) relaxation time ($\tau_1$ and $\tau_2$), and (d) the amplitude ($A_1$ and $A_2$).   Dotted line is a horizontal line for easier viewing. As $d$ increases to $d=4$ (\AA), $C_2(t^*)$ and $\tau_i$, $A_i$ all converge to a fixed value, respectively. (We also obtained consistent results for systems with 125 and 216 water molecules, please refer to Supporting Information.)}
    \label{fig:Fig5}
\end{figure*}

As shown in Figure~\ref{fig:Fig5}a, the orientation correlation function fits well with a bi-exponential decay function. The two relaxation rates ($1/\tau_1$ and $1/\tau_2$) differ; the larger relaxation time $\tau_2$ increases with the increase of $d$, while the smaller one, $\tau_1$, remains relatively unchanged with variations in $d$. In Figure~\ref{fig:Fig5}b, the values of the orientational correlation function $C_2(t)= \langle P_2(\hat u(0)\hat u(t))\rangle$ at three reference time $t^*=1, 2, 5$ (ps) are plotted. At all these times, $C(t^*)$ increases with increasing $d$ and no longer increases significantly for $d\ge 4$ (\AA).

Similar to the earlier results on HB population correlation functions and the HB reaction rates, as $d$ increases, $C_2(t^*)$ and associated relaxation times converge to a fixed value, respectively, which characterizes the decay time of the orientation relaxation process of OH bonds. From the convergence trend of the relaxation times and corresponding amplitudes in Figure~\ref{fig:Fig5}c,d, we find that at the interface with a thickness greater than 4 \AA, the OH orientation relaxation of the air-water interface is no longer different from bulk water.

Experimentally, the information on the interface molecules can be obtained from the SFG spectrum. The vibration relaxation time of the interface water molecules can be obtained based on the SFG spectrum using analysis techniques such as Singular Value Decomposition (SVD)\cite{Inoue2020}. We also defined the interface molecule orientation relaxation time from the $C_2(t)$ correlation of the interface water molecules. However, due to the arbitrariness of the specific form of defining the relaxation time, the results we obtained cannot be directly and accurately compared with the experimental results. Because of the current huge challenges in directly measuring water, we compared the data obtained from interfacial HB dynamics, interfacial OH orientation relaxation dynamics, and AIMD simulations. Our results from interfacial HB dynamics are consistent with the results from $C_2(t)$ and density profile from AIMD simulations of the air-water interface. Furthermore, the interface thickness aligns with values obtained through experimental measurements and AIMD simulations, as detailed in Table \ref{tab:interface_thickness}. Therefore, we arrive at a consistent conclusion on the issue of estimating the thickness of the air-water interface, from the perspective of HB dynamics and OH reorientation relaxation. 

\setcounter{table}{0}
\begin{table}[h]
    \caption{Air-Water Interface Thickness Obtained by Experiments and Computer Simulations.} 
    \label{tab:interface_thickness}
	\begin{tabular}{cccc}
 	\hline
		Methods & $T$ (K) & $d$ (\AA) \\
		\hline
		ellipsometry (Rayleigh)  & 293.15 & 3.0 \\
		ellipsometry (\citet{Raman1927}) & 293  & 5.0 \\
		ellipsometry (\citet{McBain1939}) & 293  & $\ge$2.26 \\
		ellipsometry (\citet{Kinosita1965}) &  293 &  7.1   \\
		X-ray reflectivity(\citet{Braslau1988}) &  298 & 3.24$\pm$ 0.05 \\
    DeePMD/MB-pol/Free OH (this work) & 300 & 5.0 \\
    BOMD/BLYP-D3/LC\&IHB (this work) & 300 & 4.0 \\
    DeePMD/MB-pol/LC\&IHB (this work) & 300 & 4.0 \\

    \hline
	\end{tabular}
\end{table}

There is one more aspect worth noting when studying the air-water interface based on the aforementioned definition of H-bonds. Some complications arise in interpreting vibrational SFG spectroscopy results~\cite{Tang2018}. An alternative perspective focusing on nonbonded OH groups better establishes a direct correlation between free OH and the 3700 cm$^{-1}$ peak, typically attributed to free OH groups~\cite{Tang2018}. Considering the dynamics of free OH groups, the thickness of the air-water interface is also estimated from simulations. Details on the methods and results can be found in Supporting Information. From the $d$-dependence of the interfacial free OH correlation function, we determined an interface thickness of approximately 5 \AA. As shown in Table \ref{tab:interface_thickness}, this value slightly deviates from the main results presented in this paper, highlighting that the measured thickness of the air-water interface varies depending on the properties of interest. Thus, to accurately determine the water interface thickness, it is essential to integrate findings from these varied perspectives.
\section{Conclusions}
\label{sec: conclusions}
In this study, we have developed a two-extremes approach to investigate the HB dynamics at the air-water interface and to determine the interface's thickness. One extreme scenario underestimates the HB breaking rate constant,  while the other overestimates it, implying that each scenario provides only a partial insight into the interfacial HB dynamics. 
Subsequently, based on DFTMD and DeePMD simulations, we have applied our approach to two distinct system properties: HB relaxation and HB reaction rate constants at the air-water interface.  Our results across both properties indicate that the predictions from both extreme scenarios converge as the thickness of the air-water interface increases to 4 \AA. Thus, we have reason to believe the thickness, which falls between these two extremes, converges at this critical value.

This work introduces an approach to complement existing ones for investigating the air-water interface from a fresh perspective. Through HB dynamics of the air-water interface, interfacial properties, such as thickness in this case, can be obtained through a method analogous to the squeeze theorem.  Beyond the scope of HB dynamics, this approach can be extended to other properties like molecular orientation distribution~\cite{RaoYi2003, GanWei2006}, free OH dynamics~\cite{Tang2018}, and SFG spectrum~\cite{Nihonyanagi2011, Nihonyanagi2015, TangFJ2020}, or other systems like solution interfacial surfaces where statistical properties of the interface and bulk phase differ significantly. Looking ahead, this study could inspire further research into ions' hydration shells by examining ions' effects through HB dynamics.

It is important to note that our current approach focuses on probing the local environment by analyzing water pairs within HB dynamics, without considering the collective behaviors of water molecules in the HB networks. However, the collective dynamics of many water molecules have been increasingly discussed in recent research, demonstrating a close relationship with observable properties such as dielectric spectroscopy and time-dependent vibrational spectroscopy.\cite{Popov2016, Fecko2003, Fecko2005, Hlzl2021, OffeiDanso2023, Malosso_2024} We hope our study will inspire further innovative ideas on related topics by incorporating considerations of the collective nature of water.

\section*{Notes}
\noindent
The authors declare no competing financial interest.

\section*{Code availability}
\noindent
The codes utilized in this study are publicly accessible on GitHub at \href{https://github.com/hg08/hb_ihb}{https://github.com/hg08/hb\_ihb}.

\begin{acknowledgments}
\noindent
This work was supported by the CSC | Chinese Government Scholarship (No. 2011113214). G.H. thanks Yuliang Jin, Rémi Khatib and Marialore Sulpizi for the useful discussions. The DFTMD simulations were conducted on the Mogon ZDV cluster in Mainz and the Cray XE6 (Hermit) at the HRLS supercomputing center in Stuttgart. The DeePMD simulations were performed on Triton, which is provided by the Aalto Science-IT project.
\end{acknowledgments}

\section*{Supporting Information}
\noindent
The Supporting Information is available free of charge at \href{https://pubs.acs.org/doi/10.1021/acs.jctc.4c00457}{https://pubs.acs.org/doi/10.1021/acs.jctc.4c00457}.
\newline

\noindent
Hydrogen bond correlation functions, HB breaking and reforming rate constants, details of Scenario 1: the LC method, rotational anisotropy decay of OH stretch at the air-water interface, computational methods including DFTMD and DeePMD simulations, results based on DFTMD simulations for systems in other sizes, results based on DeePMD simulations for systems in other sizes and perspectives derived from the correlation function of free OH groups (\href{https://pubs.acs.org/doi/suppl/10.1021/acs.jctc.4c00457/suppl_file/ct4c00457_si_001.pdf}{PDF})

\bibliography{ihb}
\end{document}